# Polar optical phonons in wurtzite spheroidal quantum dots: Theory and application to ZnO and ZnO/MgZnO nanostructures

Vladimir A. Fonoberov[*] and Alexander A. Balandin[†]

Nano-Device Laboratory, Department of Electrical Engineering
University of California–Riverside, Riverside, California 92521

Polar optical-phonon modes are derived analytically for spheroidal quantum dots with wurtzite crystal structure. The developed theory is applied to a freestanding spheroidal ZnO quantum dot and to a spheroidal ZnO quantum dot embedded into a MgZnO crystal. The wurtzite (anisotropic) quantum dots are shown to have strongly different polar optical-phonon modes in comparison with zincblende (isotropic) quantum dots. The obtained results allow one to explain and accurately predict phonon peaks in the Raman spectra of wurtzite nanocrystals, nanorods (prolate spheroids), and epitaxial quantum dots (oblate spheroids).

## I. Introduction

It is well known that in quantum dots with zincblende crystal structure there exist confined phonon modes with the frequencies equal to those of bulk transverse optical (TO) and longitudinal optical (LO) phonons and interface phonon modes with the frequencies intermediate between those of TO and LO modes [1]. Interface and confined optical phonon modes have been found for a variety of zincblende quantum dots such as spherical [1], spheroidal [2], multilayer spherical [3], and even multilayer tetrahedral [4] quantum dots. The calculated frequencies of optical phonon modes have been observed in the Raman, absorption, and photoluminescence spectra [4] of zincblende quantum dots.

Lately, quantum dots with wurtzite crystal structure, such as ZnO and GaN nanostructures, have attracted attention as very promising candidates for optoelectronic, electronic, and biological applications. It is well known that frequencies of optical phonons in small covalent nanocrystals depend on the nanocrystal size, because the nanocrystal boundary causes an uncertainty in the phonon wave vector, which results in the redshift and broadening of the phonon peak. While the above size-dependence is important for very small covalent nanocrystals, it is negligible in the ionic ZnO quantum dots with sizes larger than 4 nm. The latter is due to the fact that the polar optical phonons in ZnO are almost non-dispersive in the region of small wave vectors. Since most of the reported experimental data is for ZnO quantum dots with sizes larger than 4 nm, in the following we assume that the polar optical phonons are non-dispersive in the relevant range of the wave vectors. Due to the uniaxial anisotropy of wurtzite quantum dots, the confined and interface optical phonon modes in such quantum dots should be substantially different from those in zincblende (isotropic) quantum dots. The main difference comes from the anisotropy of the dielectric function of wurtzite crystals. In order to describe the dielectric function, we employ the Loudon model, which is widely accepted for the wurtzite nanostructures [5-7]. For example, the components of the dielectric tensor of wurtzite ZnO are [8]:

$$\varepsilon_{\perp}(\omega) = \varepsilon_{\perp}(\infty)\frac{\omega^2 - \left(\omega_{\perp,\mathrm{LO}}\right)^2}{\omega^2 - \left(\omega_{\perp,\mathrm{TO}}\right)^2}; \qquad \varepsilon_{z}(\omega) = \varepsilon_{z}(\infty)\frac{\omega^2 - \left(\omega_{z,\mathrm{LO}}\right)^2}{\omega^2 - \left(\omega_{z,\mathrm{TO}}\right)^2}, \tag{1}$$

---
[*] Electronic mail (V.A. Fonoberov): vladimir@ee.ucr.edu
[†] Electronic mail (A.A. Balandin): alexb@ee.ucr.edu

Table I. Optical dielectric constants and frequencies of polar optical phonons for two wurtzite crystals. The values for ZnO are from Ref. [8] and the values for Mg$_{0.2}$Zn$_{0.8}$O are from Ref. [9].

| wurtzite crystal | $\varepsilon_z(\infty)$ $=\varepsilon_\perp(\infty)$ | $\omega_{z,LO}$ (cm$^{-1}$) | $\omega_{z,TO}$ (cm$^{-1}$) | $\omega_{\perp1,LO}$ (cm$^{-1}$) | $\omega_{\perp1,TO}$ (cm$^{-1}$) | $\omega_{\perp2,LO}$ (cm$^{-1}$) | $\omega_{\perp2,TO}$ (cm$^{-1}$) |
|---|---|---|---|---|---|---|---|
| ZnO | 3.70 | 579 | 380 | 591 | 413 | | |
| Mg$_{0.2}$Zn$_{0.8}$O | 3.41 | 586 | 384 | 505 | 417 | 635 | 525 |

where the optical dielectric constants $\varepsilon_\perp(\infty)$ and $\varepsilon_z(\infty)$; LO phonon frequencies $\omega_{\perp,LO}$ and $\omega_{z,LO}$; and TO phonon frequencies $\omega_{\perp,TO}$ and $\omega_{z,TO}$ of bulk wurtzite ZnO are listed in Table I. The components of the dielectric tensor of some ternary wurtzite crystals such as Mg$_x$Zn$_{1-x}$O ($x < 0.33$) have more complex frequency dependence [9]:

$$\varepsilon_\perp(\omega) = \varepsilon_\perp(\infty)\frac{\omega^2 - \left(\omega_{\perp1,LO}\right)^2}{\omega^2 - \left(\omega_{\perp1,TO}\right)^2}\frac{\omega^2 - \left(\omega_{\perp2,LO}\right)^2}{\omega^2 - \left(\omega_{\perp2,TO}\right)^2}; \qquad \varepsilon_z(\omega) = \varepsilon_z(\infty)\frac{\omega^2 - \left(\omega_{z,LO}\right)^2}{\omega^2 - \left(\omega_{z,TO}\right)^2}. \qquad (2)$$

The corresponding material parameters from Eq. (2) for bulk wurtzite Mg$_{0.2}$Zn$_{0.8}$O are also listed in Table I. Zone center optical phonon frequencies of wurtzite ZnO and Mg$_{0.2}$Zn$_{0.8}$O are shown in Fig. 1. Since there are only two zone center optical phonon frequencies (one LO and one TO) in zincblende crystals, the phonon band structure of wurtzite crystals is more complex than that of zincblende crystals. It will be shown in the following that the latter fact leads to polar optical phonon modes in wurtzite quantum dots that are strongly different from those in zincblende quantum dots.

The rest of the paper is organized as follows. In Section II we present the analytical derivation of the polar optical phonon modes in spheroidal quantum dots with wurtzite crystal structure. In Sections III and IV, the developed theory is applied to a freestanding spheroidal ZnO quantum dot and to a spheroidal ZnO quantum dot embedded into a Mg$_{0.2}$Zn$_{0.8}$O crystal, correspondingly. Conclusions are given in Section V.

## II. Theory

Let us consider a spheroidal quantum dot with wurtzite crystal structure and with semi-axes $a$ and $c$. The coordinate system $(x, y, z')$ is chosen in such a way that the semi-axis $c$ is directed along the symmetry axis $z'$ of the quantum dot. The equation of the quantum dot surface is

$$\frac{x^2 + y^2}{a^2} + \frac{z'^2}{c^2} = 1 .$$

(3)

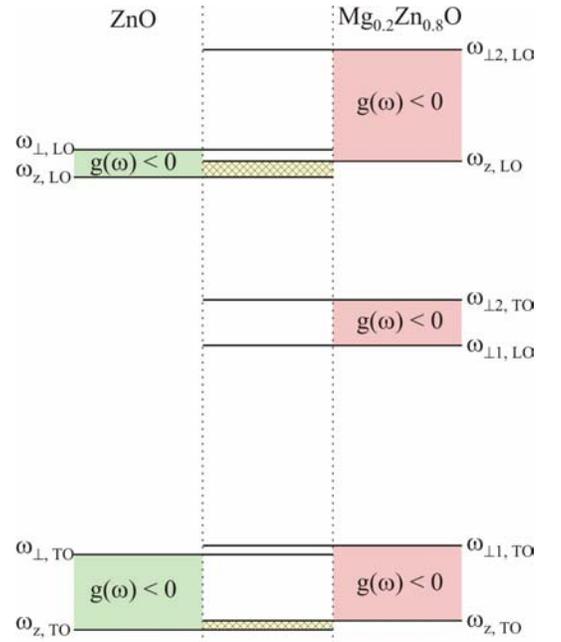

Fig. 1. (Color online). Zone center optical phonon frequencies of ZnO and Mg$_{0.2}$Zn$_{0.8}$O. Shaded regions correspond to the condition $g(\omega)<0$ [see Eq. (15)]. Cross-hatched regions correspond to the condition $g(\omega)<0$ for ZnO and $g(\omega)>0$ for Mg$_{0.2}$Zn$_{0.8}$O.

After we introduce a new coordinate $z$ such as

$$z' = \frac{c}{a} z$$

(4)

and transform the new Cartesian coordinates $(x, y, z)$ into spherical coordinates $(r, \theta, \phi)$, the Eq. (3) of the quantum dot surface becomes $r = a$. In the following derivation we assume that the quantum dot (medium $k = 1$) is embedded in a wurtzite crystal (medium $k = 2$). A freestanding quantum dot can be easily considered as a special case.

Within the framework of the dielectric-continuum approximation, the potential $V(\mathbf{r})$ of polar optical phonons satisfies the Maxwell's equation, which can be written in the coordinates $\mathbf{r} = (x, y, z)$ as

$$-\nabla \left( \hat{\varepsilon}(\omega, \mathbf{r}) \nabla V(\mathbf{r}) \right) = 0$$

(5)

with the dielectric tensor $\hat{\varepsilon}(\omega, \mathbf{r})$ defined as

$$\hat{\varepsilon}(\omega, \mathbf{r}) = \begin{pmatrix} \varepsilon_\perp(\omega, \mathbf{r}) & 0 & 0 \\ 0 & \varepsilon_\perp(\omega, \mathbf{r}) & 0 \\ 0 & 0 & \dfrac{a^2}{c^2} \varepsilon_z(\omega, \mathbf{r}) \end{pmatrix} .$$

(6)

Note that the term $a^2 / c^2$ appears in Eq. (6) due to the coordinate transformation (4). The dielectric tensor (6) is constant in both media:

$$\hat{\varepsilon}(\omega, \mathbf{r}) = \begin{cases} \hat{\varepsilon}_1(\omega), & r \le a; \\ \hat{\varepsilon}_2(\omega), & r > a, \end{cases}$$

(7)

therefore it is convenient to split Eq. (5) into separate equations for each medium:

$$-\nabla \left( \hat{\varepsilon}_k(\omega) \nabla V_k(\mathbf{r}) \right) = 0; \qquad k = 1, 2$$

(8)

and apply the corresponding boundary conditions:

$$V_1(a, \theta, \phi) = V_2(a, \theta, \phi);$$

(9)

$$D_1(a, \theta, \phi) = D_2(a, \theta, \phi),$$

(10)

where the projections of the displacement vector $\mathbf{D}$ on the outer normal $\mathbf{n}$ at the quantum dot surface can be written as

$$D_k(a, \theta, \phi) = \left( \mathbf{n}(\mathbf{r}) \hat{\varepsilon}_k(\omega) \nabla V_k(\mathbf{r}) \right) \big|_{r=a}; \qquad k = 1, 2 .$$

(11)

The phonon potential $V_1(\mathbf{r})$ that satisfies Eq. (8) and is finite everywhere inside the quantum dot can be found analytically in spheroidal coordinates $(\xi_1, \eta_1, \phi)$:

$$V_1(\mathbf{r}) = \frac{P_l^m(\xi_1)}{P_l^m(\xi_1^{(0)})} P_l^m(\eta_1) e^{im\phi} .$$

(12)

Analogously, the phonon potential $V_2(\mathbf{r})$ that satisfies Eq. (8) and vanishes far away from the quantum dot can be found analytically in spheroidal coordinates $(\xi_2, \eta_2, \phi)$:

$$V_2(\mathbf{r}) = \frac{Q_l^m(\xi_2)}{Q_l^m(\xi_2^{(0)})} P_l^m(\eta_2) e^{im\phi} .$$

(13)

In Eqs. (12) and (13), $P_l^m$ and $Q_l^m$ are associated Legendre functions of the first and second kinds, respectively; the integers $l$ ($l \geq 0$) and $m$ ($|m| \leq l$) are quantum numbers of the phonon mode. The spheroidal coordinates $(\xi_k, \eta_k)$ are related to the spherical coordinates $(r, \theta)$ as

$$\begin{cases} r \sin\theta = a\sqrt{\left(\dfrac{1}{g_k(\omega)}-1\right)\left(\xi_k^2 - 1\right)}\sqrt{1-\eta_k^2}, \\ r\cos\theta = a\sqrt{1-g_k(\omega)}\ \xi_k\ \eta_k, \end{cases} \tag{14}$$

where $k = 1, 2$ and

$$g_k(\omega) = \frac{a^2}{c^2}\frac{\varepsilon_z^{(k)}(\omega)}{\varepsilon_\perp^{(k)}(\omega)}. \tag{15}$$

The range of the spheroidal coordinate $\eta_k$ is $-1 \leq \eta_k \leq 1$. Depending on the value of the function (15), the spheroidal coordinate $\xi_k$ can have the following range:

$$\begin{aligned} &0 < \xi_k < 1 \quad \text{if} \quad g_k(\omega) < 0; \\ &\xi_k > 1 \qquad \text{if} \quad 0 < g_k(\omega) < 1; \\ &\mathrm{i}\xi_k > 0 \qquad \text{if} \quad g_k(\omega) > 1. \end{aligned} \tag{16}$$

According to Eq. (14), the quantum dot surface $r = a$ is defined in the spheroidal coordinates as

$$\begin{cases} \xi_k = \xi_k^{(0)} \equiv 1/\sqrt{1-g_k(\omega)}, \\ \eta_k = \cos\theta. \end{cases} \tag{17}$$

Therefore, the part of the phonon potential $V_1(\mathbf{r})$ defined by Eq. (12) and the part of the phonon potential $V_2(\mathbf{r})$ defined by Eq. (13) coincide at the quantum dot surface. Thus, the first boundary condition, given by Eq. (9), is satisfied.

Now, let us find the normal component of the displacement vector $\mathbf{D}$ at the quantum dot surface. According to Eq. (11),

$$D_k(a,\theta,\phi) = \varepsilon_\perp^{(k)}(\omega)\left[\left(g_k(\omega)\cos^2\theta + \sin^2\theta\right)\frac{\partial V_k}{\partial r}\bigg|_{r=a} + \frac{1-g_k(\omega)}{a}\sin\theta\cos\theta\frac{\partial V_k}{\partial \theta}\bigg|_{r=a}\right]. \tag{18}$$

Using relation (14) between the coordinates $(\xi_k, \eta_k)$ and $(r, \theta)$, we can calculate each of the two partial derivatives from Eq. (18):

$$\frac{\partial V_k}{\partial r}\bigg|_{r=a} = \frac{1}{a\left(g_k(\omega)\cos^2\theta + \sin^2\theta\right)}\left[\frac{g_k(\omega)}{\sqrt{1-g_k(\omega)}}\frac{\partial V_k}{\partial \xi_k}\bigg|_{\substack{\xi_k=\xi_k^{(0)}\\ \eta_k=\cos\theta}} + \cos\theta\sin^2\theta\left(1-g_k(\omega)\right)\frac{\partial V_k}{\partial \eta_k}\bigg|_{\substack{\xi_k=\xi_k^{(0)}\\ \eta_k=\cos\theta}}\right], \tag{19}$$

$$\frac{\partial V_k}{\partial \theta}\bigg|_{r=a} = -\sin\theta\frac{\partial V_k}{\partial \eta_k}\bigg|_{\substack{\xi_k=\xi_k^{(0)}\\ \eta_k=\cos\theta}}. \tag{20}$$

Substituting Eqs. (19) and (20) into Eq. (18), one obtains a simple formula:

$$D_k(a,\theta,\phi) = \frac{\varepsilon_\perp^{(k)}(\omega)g_k(\omega)}{a\sqrt{1-g_k(\omega)}}\frac{\partial V_k}{\partial \xi_k}\bigg|_{\substack{\xi_k=\xi_k^{(0)}\\ \eta_k=\cos\theta}}. \tag{21}$$

Finally, using the explicit form of the phonon potentials (12) and (13) as well as Eqs. (15) and (17), one can rewrite Eq. (21) as

$$D_1(a,\theta,\phi) = \frac{a}{c^2} \frac{\varepsilon_z^{(1)}(\omega)}{\sqrt{1-g_1(\omega)}} \frac{d\ln P_l^m(\xi_1)}{d\xi_1}\Bigg|_{\xi_1=\xi_1^{(0)}} P_l^m(\cos\theta)\, e^{im\phi}\,; \qquad (22)$$

$$D_2(a,\theta,\phi) = \frac{a}{c^2} \frac{\varepsilon_z^{(2)}(\omega)}{\sqrt{1-g_2(\omega)}} \frac{d\ln Q_l^m(\xi_2)}{d\xi_2}\Bigg|_{\xi_2=\xi_2^{(0)}} P_l^m(\cos\theta)\, e^{im\phi}\,. \qquad (23)$$

Substituting Eqs. (22) and (23) into the second boundary condition (10), one can see that it is satisfied only when the following equality is true

$$\varepsilon_z^{(1)}(\omega)\left(\xi\frac{d\ln P_l^m(\xi)}{d\xi}\right)\Bigg|_{\xi=1/\sqrt{1-g_1(\omega)}} = \varepsilon_z^{(2)}(\omega)\left(\xi\frac{d\ln Q_l^m(\xi)}{d\xi}\right)\Bigg|_{\xi=1/\sqrt{1-g_2(\omega)}}. \qquad (24)$$

Thus, we have obtained the equation that defines the spectrum of polar optical phonons in a wurtzite spheroidal quantum dot embedded in a wurtzite crystal. It should be pointed out, that for a spheroidal quantum dot with zincblende crystal structure $\varepsilon_\perp^{(k)}(\omega) = \varepsilon_z^{(k)}(\omega) \equiv \varepsilon^{(k)}(\omega)$ and Eq. (24) reduces to the one obtained in Ref. [2]. The fact that the spectrum of polar optical phonons does not depend on the absolute size of a quantum dot [1,2] is also seen from Eq. (24).

The case of a freestanding quantum dot is no less important for practical applications. In this case the dielectric tensor of the exterior medium is a constant $\varepsilon_D \equiv \varepsilon_z^{(2)}(\omega) = \varepsilon_\perp^{(2)}(\omega)$. Therefore, using the explicit form of associated Legendre polynomials $P_l^m$ and omitting the upper index "(1)" in the components of the dielectric tensor of the quantum dot, we can represent Eq. (24) in the following convenient form:

$$\sum_{n=0}^{\left\lfloor\frac{l-|m|}{2}\right\rfloor}\left[\frac{c^2}{a^2}\frac{\varepsilon_\perp(\omega)}{\varepsilon_D}|m| + \frac{\varepsilon_z(\omega)}{\varepsilon_D}\big(l-|m|-2n\big) - f_l^{|m|}\left(\frac{a}{c}\right)\right]$$
$$\times\binom{l-|m|}{2n}\frac{(2n-1)!!(2l-2n-1)!!}{(2l-1)!!}\left[\frac{a^2}{c^2}\frac{\varepsilon_z(\omega)}{\varepsilon_\perp(\omega)}-1\right]^n = 0, \qquad (25)$$

where

$$f_l^m(\alpha) = \xi\frac{d\ln Q_l^m(\xi)}{d\xi}\Bigg|_{\xi=1/\sqrt{1-\alpha^2}}. \qquad (26)$$

It can be shown that the function $f_l^m(\alpha)$ increases monotonely from $-\infty$ to 0 when $\alpha$ increases from 0 to $\infty$. As seen from Eq. (25), there are no phonon modes with $l=0$ and all phonon frequencies with $m\neq 0$ are twice degenerate with respect to the sign of $m$. For a spherical ($\alpha=1$) freestanding quantum dot one has to take the limit $\xi\to\infty$ in Eq. (26), what results in $f_l^m(1) = -(l+1)$. Thus, in the case of a zincblende spherical quantum dot $[\varepsilon_\perp(\omega) = \varepsilon_z(\omega) \equiv \varepsilon(\omega)$; $a = c]$, Eq. (25) gives the well-known equation $\varepsilon(\omega)/\varepsilon_D = -1 - 1/l$ derived in Ref. [1].

## III.   Freestanding ZnO quantum dots

In this Section we consider freestanding spheroidal ZnO quantum dots and examine the phonon modes with quantum numbers $l$ = 1, 2, 3, 4 and $m$ = 0, 1. The components of the dielectric tensor of wurtzite ZnO are given by Eq. (1). The exterior medium is considered to be air with $\varepsilon_D = 1$. Fig. 2 shows the spectrum of polar optical phonons with $m = 0$ and Fig. 3 shows the spectrum of polar optical phonons with $m = 1$. The frequencies with even $l$ are plotted with solid curves while the frequencies with odd $l$ are plotted with dashed curves. The frequencies in Figs. 2 and 3 are found as solutions of Eq. (25) and are plotted as a function of the ratio of the spheroidal semi-axes $a$ and $c$. Thus, in the leftmost part of the plots we have the phonon spectrum for a spheroid degenerated into a vertical line segment; father to the right we have the spectrum for prolate spheroids; in the central part of the plots we have the phonon spectrum for a sphere; farther on we have the spectrum for oblate spheroids; and in the rightmost part of the plots we have the phonon spectrum for a spheroid degenerated into a horizontal flat disk.

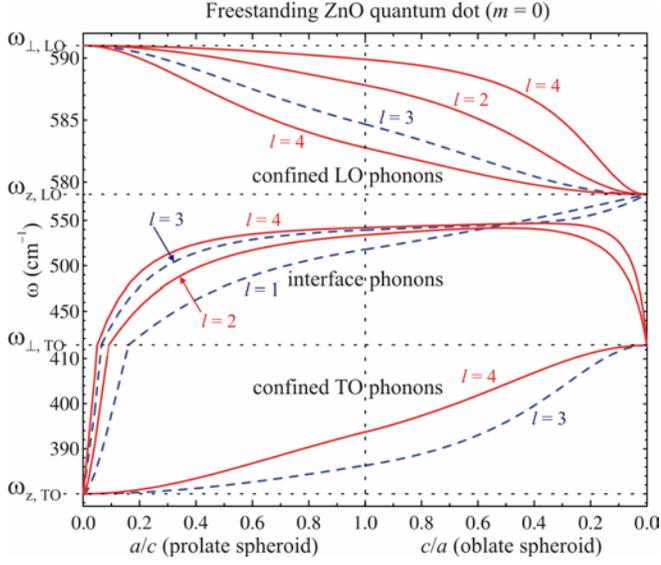

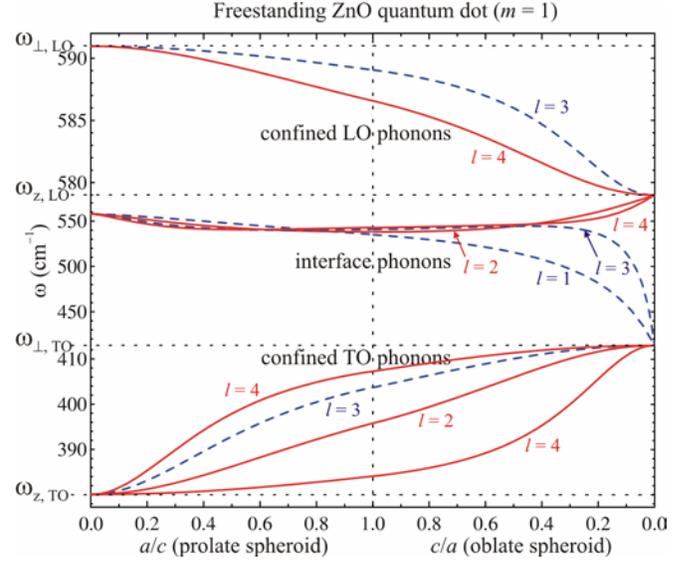

Fig. 2. (Color online). Frequencies of polar optical phonons with $l = 1, 2, 3, 4$ and $m = 0$ for a freestanding spheroidal ZnO quantum dot as a function of the ratio of spheroidal semi-axes. Solid curves correspond to phonons with even $l$ and dashed curves correspond to phonons with odd $l$. Frequency scale is different for confined TO, interface, and confined LO phonons.

Fig. 3. (Color online). The same as Fig. 2 but for polar optical phonons with $m = 1$.

The calculated spectrum of phonons in the freestanding ZnO quantum dots can be divided into three regions: confined TO phonons ($\omega_{z,\text{TO}} < \omega < \omega_{\perp,\text{TO}}$), interface phonons ($\omega_{\perp,\text{TO}} < \omega < \omega_{z,\text{LO}}$), and confined LO phonons ($\omega_{z,\text{LO}} < \omega < \omega_{\perp,\text{LO}}$). The above division into confined and interface phonons is based on the sign of the function $g(\omega)$ [see Eq. (15)]. We call the phonons with eigenfrequency $\omega$ interface phonons if $g(\omega) > 0$ and confined phonons if $g(\omega) < 0$. To justify the classification of phonon modes as interface and confined ones based on the sign of the function $g_1(\omega)$, let us consider the phonon potential (12) inside the quantum dot. If $g_1(\omega) < 0$ then, according to Eq. (16), $0 < \xi_1 < 1$; therefore $P_l^m(\xi_1)$ is an oscillatory function of $\xi_1$ and the phonon potential (12) is mainly confined inside the quantum dot. On the contrary, if $g_1(\omega) > 0$ then, according to Eq. (16), $\xi_1 > 1$ or $i\xi_1 > 0$; therefore $P_l^m(\xi_1)$ increases monotonely with $\xi_1$ as $\xi_1^l$ reaching the maximum at the quantum dot surface together with the phonon potential (12). Note that vertical frequency scale in Figs. 2 and 3 is different for confined TO, interface, and confined LO phonons. The true scale is shown in Fig. 1.

Analyzing Eq. (25), one can find that for each pair $(l, m)$ there is one interface optical phonon and $l - |m|$ confined optical phonons for $m \neq 0$ ($l - 1$ for $m = 0$). Therefore, we can see four interface phonons and six confined phonons for both $m = 0$ and $m = 1$ in Figs. 2 and 3. However, one can see that there are four confined LO phonons with $m = 0$ and only two confined LO phonons with $m = 1$. On the contrary, there are only two confined TO phonons with $m = 0$ and four confined TO phonons with $m = 1$ in Figs. 2 and 3.

When the shape of the spheroidal quantum dot changes from the vertical line segment to the horizontal flat disk, the frequencies of all confined LO phonons decrease from $\omega_{\perp,\text{LO}}$ to $\omega_{z,\text{LO}}$. At the same time the frequencies of all confined TO phonons increase from $\omega_{z,\text{TO}}$ to $\omega_{\perp,\text{TO}}$. It is also seen from Figs. 2 and 3 that for very small ratios $a/c$, what is the case of so-called quantum rods, the interface phonons with $m = 0$ become confined TO phonons, while the frequencies of all interface phonons with $m = 1$ degenerate into a single frequency. When the shape of the spheroidal quantum dot changes from the vertical line segment to the horizontal flat disk, the frequencies of interface phonons with odd $l$ and $m = 0$ increase from $\omega_{z,\text{TO}}$ to $\omega_{z,\text{LO}}$, while the frequencies of interface phonons with even $l$ and $m = 0$ increase for prolate spheroids starting from $\omega_{z,\text{TO}}$, like for the phonons with odd $l$, but they farther decrease up to $\omega_{\perp,\text{TO}}$ for oblate spheroids. On the contrary, when the shape of the spheroidal quantum dot changes from the vertical line segment to the horizontal flat disk, the frequencies of interface phonons with odd $l$ and $m = 1$ decrease from a single interface frequency to $\omega_{\perp,\text{TO}}$, while the frequencies of interface phonons with even $l$ and $m = 1$ decrease for prolate spheroids starting from a single frequency, like for the phonons with odd $l$, but they farther increase up to $\omega_{z,\text{LO}}$ for oblate spheroids.

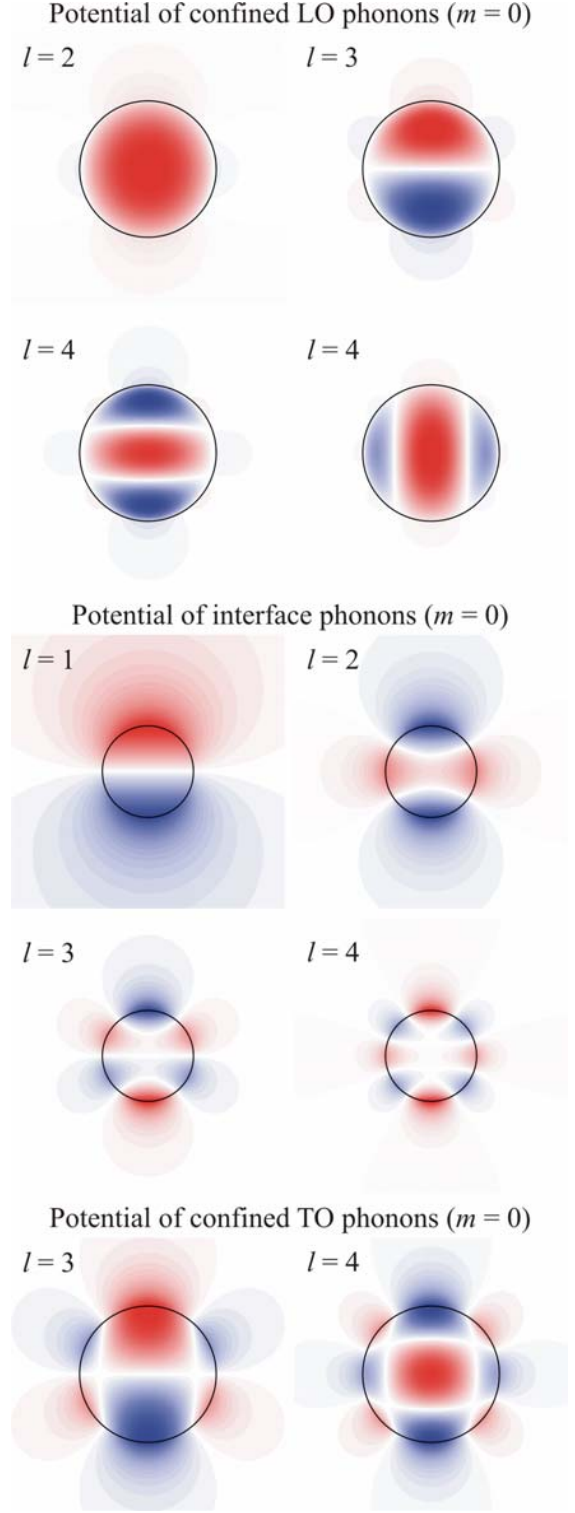

Fig. 4. (Color online). Cross-sections of phonon potentials corresponding to polar optical phonon modes with $l = 1, 2, 3, 4$ and $m = 0$ for the freestanding spherical ZnO quantum dot. Z-axis is directed vertically. In the color version blue and red colors denote negative and positive values of phonon potentials, correspondingly. Black circle represents the quantum dot surface.

In the rest of this Section we study phonon potentials corresponding to the polar optical phonon modes with $l = 1, 2, 3, 4$ and $m = 0$. In Fig. 4 we present the phonon potentials for a spherical freestanding ZnO quantum dot. The phonon potentials for quantum dots with arbitrary spheroidal shapes can be found analogously using Eqs. (12), (13) and the coordinate transformation (4). As seen from Fig. 4, the confined LO phonons are, indeed, confined inside the quantum dot. However, unlike confined phonons in zincblende quantum dots, confined phonons in wurtzite quantum dots slightly penetrate into the exterior medium. Potential of interface phonon modes is, indeed, localized near the surface of the wurtzite quantum dot. While there are no confined TO phonons in zincblende quantum dots, they appear in wurtzite quantum dots. It is seen from Fig. 4 that confined TO phonons are, indeed, localized mainly inside the quantum dot. However, they penetrate into the exterior medium much stronger than confined LO phonons.

Using the theory of excitonic states in wurtzite quantum dots [10-11], it can be shown that the dominant component of the wave function of the exciton ground state in spheroidal ZnO quantum dots is symmetric with respect to the rotations around the $z$-axis or reflection in the $xy$-plane. Therefore, the selection rules for the polar optical phonon modes observed in the resonant Raman spectra of ZnO quantum dots are $m = 0$ and $l = 2, 4, 6, \ldots$ The phonon modes with higher symmetry (smaller quantum number $l$) are more likely to be observed in the Raman spectra. It is seen from Fig. 4, that the confined LO phonon mode with $l = 2, m = 0$ and the confined TO mode with $l = 4, m = 0$ are the confined modes with the highest symmetry among the confined LO and TO phonon modes, correspondingly. Therefore, they should give the main contribution to the resonant Raman spectrum of spheroidal ZnO quantum dots.

In fact, the above conclusion has an experimental confirmation. In the resonant Raman spectrum of spherical ZnO quantum dots with diameter 8.5 nm from Ref. [12], the main Raman peak in the region of LO phonons has the frequency 588 cm$^{-1}$ and the main Raman peak in the region of TO phonons has the frequency 393 cm$^{-1}$. In accordance with Fig. 2, our calculations give the frequency 587.8 cm$^{-1}$ of the confined LO phonon mode with $l = 2, m = 0$ and the frequency 393.7 cm$^{-1}$ of the confined TO phonon mode with $l = 4, m = 0$. This excellent agreement of the experimental and calculated frequencies allows one to predict the main peaks in the LO and TO regions of a Raman spectra of spheroidal ZnO quantum dots using the corresponding curves from Fig. 2.

## IV.    ZnO/MgZnO quantum dots

In this Section we consider spheroidal ZnO quantum dots embedded into a $Mg_{0.2}Zn_{0.8}O$ crystal. The components of the dielectric tensors of wurtzite ZnO and $Mg_{0.2}Zn_{0.8}O$ are given by Eqs. (1) and (2), correspondingly. The relative position of optical phonon bands of wurtzite ZnO and $Mg_{0.2}Zn_{0.8}O$ is shown in Fig. 1. It is seen from Eq. (15) that $g_1(\omega) < 0$ inside the shaded region corresponding to ZnO in Fig. 1 and $g_2(\omega) < 0$ inside the shaded region corresponding to $Mg_{0.2}Zn_{0.8}O$. As it has been shown in Section III, the frequency region where $g_1(\omega) < 0$ corresponds to confined phonons in a freestanding spheroidal ZnO quantum dot. However, there can be no confined phonons in the host $Mg_{0.2}Zn_{0.8}O$ crystal. Indeed, there are no physical solutions of Eq. (24) when $g_2(\omega) < 0$. The solutions of Eq. (24) are nonphysical in this case, because the spheroidal coordinates $(\xi_2, \eta_2)$ defined by Eq. (14) cannot cover the entire space outside the quantum dot. If we allow the spheroidal coordinates $(\xi_2, \eta_2)$ to be complex, then the phonon potential outside the quantum dot becomes complex and diverges logarithmically when $\xi_2 = 1$; the latter is clearly nonphysical. It can be also shown that Eq. (24) does not have any solutions when $g_1(\omega) > 0$ and $g_2(\omega) > 0$. Therefore, the only case when Eq. (24) allows for physical solutions is

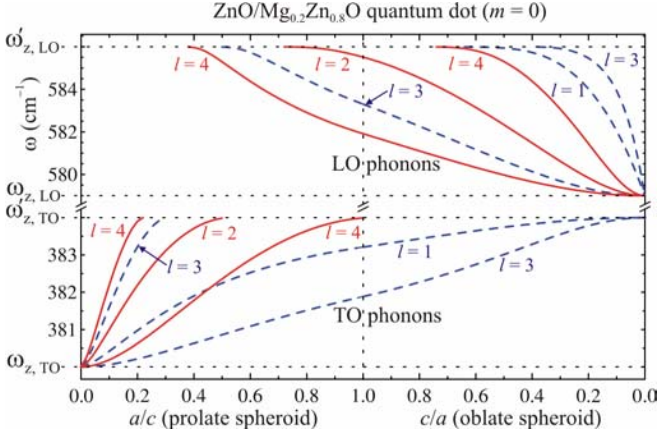

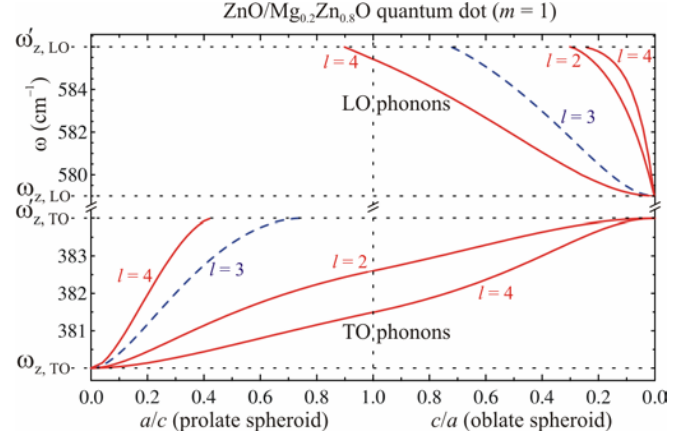

Fig. 5. (Color online). Frequencies of polar optical phonons with $l = 1, 2, 3, 4$ and $m = 0$ for a spheroidal ZnO/Mg$_{0.2}$Zn$_{0.8}$O quantum dot as a function of the ratio of spheroidal semi-axes. Solid curves correspond to phonons with even $l$ and dashed curves correspond to phonons with odd $l$. Frequency scale is different for TO and LO phonons. Frequencies $\omega_{z,\text{TO}}$ and $\omega_{z,\text{LO}}$ correspond to ZnO and frequencies $\omega'_{z,\text{TO}}$ and $\omega'_{z,\text{LO}}$ correspond to Mg$_{0.2}$Zn$_{0.8}$O.

Fig. 6. (Color online). The same as Fig. 5 but for polar optical phonons with $m = 1$.

$g_1(\omega) < 0$ and $g_2(\omega) > 0$. The frequency regions that satisfy the latter condition are cross-hatched in Fig. 1. There are two such regions: $\omega_{z,\text{TO}}^{(1)} < \omega < \omega_{z,\text{TO}}^{(2)}$ and $\omega_{z,\text{LO}}^{(1)} < \omega < \omega_{z,\text{LO}}^{(2)}$, which are further called the regions of TO and LO phonons, respectively.

Let us now examine the LO and TO phonon modes with quantum numbers $l = 1, 2, 3, 4$ and $m = 0, 1$. Fig. 5 shows the spectrum of polar optical phonons with $m = 0$ and Fig. 6 shows the spectrum of polar optical phonons with $m = 1$. The frequencies with even $l$ are plotted with solid curves while the frequencies with odd $l$ are plotted with dashed curves. The frequencies in Figs. 5 and 6 are found as solutions of Eq. (24) and are plotted as a function of the ratio of the spheroidal semi-axes $a$ and $c$, similarly to Figs. 2 and 3 for the freestanding spheroidal ZnO quantum dot. Note that vertical frequency scale in Figs. 5 and 6 is different for TO phonons and LO phonons. The true scale is shown in Fig. 1.

Comparing Fig. 5 with Fig. 2 and Fig. 6 with Fig. 3 we can see the similarities and distinctions in the phonon spectra of the ZnO quantum dot embedded into the Mg$_{0.2}$Zn$_{0.8}$O crystal and that of the freestanding ZnO quantum dot. For a small ratio $a/c$ we have the same number of TO phonon modes with the frequencies originating from $\omega_{z,\text{TO}}^{(1)}$ for the embedded and freestanding ZnO quantum dots. With the increase of the ratio $a/c$ the frequencies of TO phonons increase for both embedded and freestanding ZnO quantum dots, but the number of TO phonon modes gradually decreases in the embedded ZnO quantum dot. When $a/c \to \infty$ there are only two phonon modes with odd $l$ are left for $m = 0$ and two phonon modes with even $l$ are left for $m = 1$. The frequencies of these phonon modes increase up to $\omega_{z,\text{TO}}^{(2)}$ when $a/c \to \infty$. However, for this small ratio $c/a$ we have the same number of LO phonon modes with the frequencies originating from $\omega_{z,\text{LO}}^{(1)}$ for the embedded and freestanding ZnO quantum dots. With the increase of the ratio $c/a$ the frequencies of all LO phonons increase for the embedded ZnO quantum dot and the number of such phonons gradually decreases. When $c/a \to \infty$ there are no phonons left for the embedded ZnO quantum dot. At the same time for the freestanding ZnO quantum dot, with the increase of the ratio $c/a$ the frequencies of two LO phonons with odd $l$ and $m = 0$ and two LO phonons with even $l$ and $m = 1$ decrease and go into the region of interface phonons.

It is seen from the previous paragraph that for the ZnO quantum dot with a small ratio $c/a$ embedded into the $Mg_{0.2}Zn_{0.8}O$ crystal the two LO and two TO phonon modes with odd $l$ and $m = 0$ and with even $l$ and $m = 1$ may correspond to interface phonons. To check this hypothesis, we further study phonon potentials corresponding to the polar optical phonon modes with $l = 1, 2, 3, 4$ and $m = 0$. In Fig. 7 we present the phonon potentials for the spheroidal ZnO quantum dot with the ratio $c/a = 1/4$ embedded into the $Mg_{0.2}Zn_{0.8}O$ crystal. The considered ratio $c/a = 1/4$ of the spheroidal semi-axes is a reasonable value for epitaxial ZnO/ $Mg_{0.2}Zn_{0.8}O$ quantum dots. It is seen in Fig. 7 that the LO phonon with $l = 1$, one of the LO phonons with $l = 3$, and all two TO phonons are, indeed, interface phonons, since they achieve their maximal and minimal values at the surface of the ZnO quantum dot. It is interesting that the potential of interface TO phonons is strongly extended along the $z$-axis, while the potential of interface LO phonons is extended in the $xy$-plane. All other LO phonons in Fig. 7 are confined. The

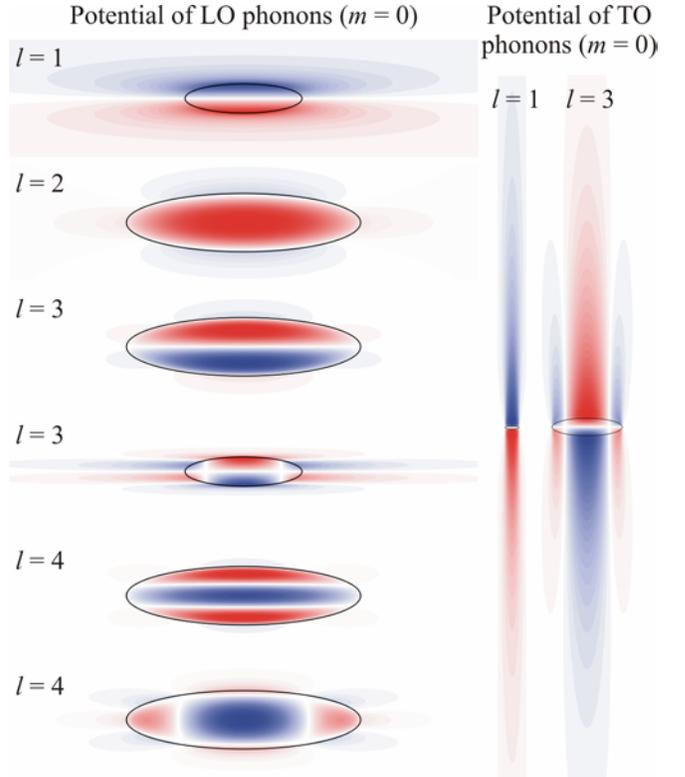

Potential of LO phonons ($m = 0$)

Potential of TO phonons ($m = 0$)

$l = 1$

$l = 2$

$l = 3$

$l = 3$

$l = 4$

$l = 4$

$l = 1$   $l = 3$

Fig. 7. (Color online). Cross-sections of phonon potentials corresponding to polar optical phonon modes with $l = 1, 2, 3, 4$ and $m = 0$ for the oblate spheroidal ZnO/$Mg_{0.2}Zn_{0.8}O$ quantum dot with aspect ratio 1/4. $Z$-axis is directed vertically. In the color version blue and red colors denote negative and positive values of phonon potentials, correspondingly. Black ellipse represents the quantum dot surface.

most symmetrical phonon mode is, again, the one with $l = 2$ and $m = 0$. Therefore, it should give the main contribution to the Raman spectrum of oblate spheroidal ZnO quantum dots embedded into the $Mg_{0.2}Zn_{0.8}O$ crystal. Unlike for freestanding ZnO quantum dots, no pronounced TO phonon peaks are expected for the embedded ZnO quantum dots.

## V. Conclusions

In conclusion, we have derived analytically interface and confined polar optical phonon modes for spheroidal quantum dots with wurtzite crystal structure. The developed theory has been applied to study phonon frequencies and potentials as a function of the ratio of spheroidal semi-axes for freestanding spheroidal ZnO quantum dots and spheroidal ZnO quantum dots embedded into the $Mg_{0.2}Zn_{0.8}O$ crystal. It has been shown that wurtzite quantum dots have strongly different polar optical phonon modes in comparison with zincblende quantum dots. While the frequency of confined polar optical phonons in zincblende quantum dots is equal to that of the bulk crystal phonons, the confined polar optical phonons in wurtzite quantum dots have a discrete spectrum of frequencies different from those of the bulk crystal. The positions of polar optical phonon lines observed in the resonant Raman spectra of spherical wurtzite ZnO quantum dots have been explained quantitatively. The obtained theoretical results allow one to explain and accurately predict phonon peaks in the Raman spectra not only for wurtzite ZnO nanocrystals, nanorods, and epitaxial $ZnO/Mg_{0.2}Zn_{0.8}O$ quantum dots, but also for any wurtzite spheroidal quantum dot either freestanding or embedded into another crystal.

**Acknowledgement**

This work was supported in part by the Microelectronics Advanced Research Corporation (MARCO) and its Focus Center on Functional Engineered Nano Architectonics (FENA), NSF-NATO 2003 award to V.A.F., ONR Young Investigator Award to A.A.B., and DMEA/DARPA CNID program A01809-23103-44.

**References**

1. R. Englman and R. Ruppin, Phys. Rev. Lett. **16**, 898 (1966).
2. P. A. Knipp and T. L. Reinecke, Phys. Rev. B **46**, 10310 (1992).
3. S. N. Klimin, E. P. Pokatilov, and V. M. Fomin, Phys. Stat. Sol. B **184**, 373 (1994).
4. V. A. Fonoberov, E. P. Pokatilov, V. M. Fomin, and J. T. Devreese, Phys. Rev. Lett. **92**, 127402 (2004).
5. M. A. Stroscio and M. Dutta, *Phonons in Nanostructures* (Cambridge University Press, Cambridge, 2001).
6. C. Chen, M. Dutta, and M. A. Stroscio, Phys. Rev. B **70**, 075316 (2004).
7. C. Chen, M. Dutta, and M. A. Stroscio, J. Appl. Phys. **96**, 2049 (2004).
8. C. A. Arguello, D. L. Rousseau, and S. P. S. Porto, Phys. Rev. **181**, 1351 (1969).
9. C. Bundesmann, M. Schubert, D. Spemann, T. Butz, M. Lorenz, E. M. Kaidashev, M. Grundmann, N. Ashkenov, H. Neumann, and G. Wagner, Appl. Phys. Lett. **81**, 2376 (2002).
10. V. A. Fonoberov and A. A. Balandin, J. Appl. Phys. **94**, 7178 (2003); V. A. Fonoberov and A. A. Balandin, J. Vac. Sci. Technol. B **22**, 2190 (2004).
11. V. A. Fonoberov and A. A. Balandin, to be published.
12. M. Rajalakshmi, A. K. Arora, B. S. Bendre, and S. Mahamuni, J. Appl. Phys. **87**, 2445 (2000).